\documentclass[prl,twocolumn]{revtex4}
\usepackage{graphicx}
\usepackage{dcolumn}
\usepackage{amsmath,amsbsy,amssymb,graphics}
\usepackage{bm}

\begin{document}

\preprint{APS/123-QED}

\title{
Diagonal composite order in two-channel Kondo lattice
}

\author{Shintaro Hoshino, Junya Otsuki and Yoshio Kuramoto}
\affiliation{Department of Physics, Tohoku University, Sendai 980-8578, Japan}

\date{\today}

\begin{abstract}
A novel type of symmetry breaking is reported
for the two-channel Kondo lattice where 
conduction electrons have spin and orbital (channel) degrees of freedom.
Using the continuous-time quantum Monte Carlo and the dynamical mean-field theory,
a spontaneous breaking of the orbital symmetry is observed.  The tiny breakdown of orbital occupation number, however, vanishes if the 
conduction electrons have the particle-hole symmetry.   The proper order parameter instead
is identified as a composite quantity representing the orbital-selective Kondo effect.  
The single-particle spectrum of the selected orbital shows insulating property, 
while the other orbital behaves as Fermi liquid.
This composite 
order 
is the first example of odd-frequency order other than off-diagonal order (superconductivity), and
is a candidate of hidden order in $f$-electron systems.
\end{abstract}

\pacs{Valid PACS appear here}
\maketitle

\newcommand{\diff}{\mathrm{d}}
\newcommand{\imag}{\mathrm{Im}\,}
\newcommand{\real}{\mathrm{Re}\,}
\newcommand{\trace}{\mathrm{Tr}\,}
\newcommand{\imu}{\mathrm{i}}


Localized $f$ electrons with even number per site 
can have a non-Kramers doublet ground state.
If these $f$ electrons couple with 
conduction electrons with orbital degeneracy,
a two-channel Kondo system can be realized  \cite{cox98}.
Recently two-channel Kondo systems have been attracting renewed attention, partly due to discovery of new $\Gamma_3$ doublet compounds such as PrIr$_2$Zn$_{20}$ and PrV$_2$Al$_{20}$  \cite{onimaru10,onimaru11,sakai11}.
Mysterious behaviors observed in PrAg$_2$In  \cite{yatskar97,suzuki06} and URu$_2$Si$_2$  \cite{kuramoto09} and their impurity counterparts such as
Pr$_x$La$_{1-x}$Ag$_2$In  \cite{kawae05} and U$_x$Th$_{1-x}$Ru$_2$Si$_2$ \cite{amitsuka94} have remained long-standing puzzles, and 
the two-channel Kondo effect has been suspected as an origin. 

The two-channel Kondo impurity has the residual entropy 
in the ground state  \cite{cox98}.
Then the two-channel Kondo lattice (2ch KL)
should undergo some phase transition to 
reach the ground state from high-temperature phase.
Possible orderings in the 2ch KL have been studied
by the dynamical mean-field theory (DMFT) 
 \cite{jarrell96,jarrell97,nourafkan08}, and by other methods \cite{schauerte05}.
Among possible ground states, the simplest is a magnetically ordered state \cite{jarrell96}.
Superconductivity with odd frequency pairing has also been discussed for the 2ch KL \cite{jarrell97}
and a related model called the SU(2)$\times$SU(2) Anderson lattice \cite{anders02}.
On the other hand, an exact diagonalization study, using a small discretized reservoir in the DMFT, has found a channel symmetry breaking \cite{nourafkan08}.
In the effective impurity problem of the DMFT, 
a minute energy scale appears associated with the Kondo effect. 
Then highly accurate treatment of the reservoir is required for numerical approach. 
Recently, a Monte-Carlo scheme using the continuous imaginary time (CT-QMC) \cite{gull11}
has proven powerful enough in solving the effective impurity problem at finite temperature \cite{otsuki07,hoshino10}.

In this letter, by combining DMFT and CT-QMC, we report on a new kind of
electronic order in the 2ch KL at and near half-filling of two equivalent conduction bands. 
Although the symmetry of channel occupation is broken, the corresponding order parameter is tiny and vanishes in the case with the particle-hole symmetry.
The proper order parameter is a composite quantity that becomes order of unity.
To our knowledge, this is the first example of the composite order parameter other than off-diagonal order (superconductivity) \cite{berezinskii74,emery92,balatsky93,schrieffer94,abrahams95,coleman95,anders02}.
It is shown that the present composite order can be interpreted also as an odd-frequency order.

The simplest description of periodic non-Kramers doublets coupled with conduction electrons is given by the 
2ch KL
\cite{jarrell96,jarrell97,schauerte05,nourafkan08}:
\begin{align}
{\cal H} = 
\sum _{\bm{k}\alpha \sigma} ( \varepsilon _{\bm{k}} -\mu ) c_{\bm{k}\alpha \sigma}^\dagger c_{\bm{k}\alpha \sigma}
+ J \sum_{i\alpha} \bm{S}_i \cdot \bm{s}_{{\rm c}i\alpha }
, \label{eqn_2ch_KLM}
\end{align}
where $c_{\bm{k}\alpha \sigma}$ is annihilation operator of the conduction electron with momentum $\bm k$, channel $\alpha =1,2$ and pseudo-spin $\sigma = \uparrow, \downarrow$.
The localized pseudo-spin at site $i$ is described by the spin 1/2 operator $\bm{S}_i$, while 
$\bm{s}_{{\rm c} i \alpha}$
represents the conduction pseudo-spin of channel $\alpha$ at site $i$.
For the moment, we call the pseudo-spin simply as `spin'.  Near the end of the letter, we return to physical consequence of the non-Kramers doublet described by the pseudo-spin.

In our DMFT, the semi-circular density of states is taken for conduction electrons defined by
$\rho_0 (\varepsilon) = (2/\pi) \sqrt{1 - (\varepsilon / D) ^2}$ with the nesting property $\varepsilon_{\bm{k}+\bm{Q}} = - \varepsilon_{\bm{k}}$ where $\bm{Q}$ is a staggered ordering vector.
We take $D=1$ as a unit of energy.
At half filling of conduction bands, the system has a particle-hole symmetry.
We have employed the Pad\'{e} approximation for analytic continuation to real frequencies.

\begin{figure}
\begin{center}
\includegraphics[width=75mm]{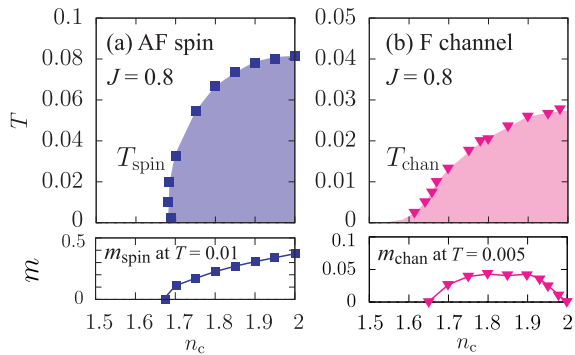}
\caption{
(color online)
Phase diagram of the 2ch KL near half filling for (a) AF-spin and (b) F-channel ordered phases.
In (b), we neglect the AF-spin order for all $n_{\rm c}$.
The lower panels show spin and channel moments close to the ground state, as defined by eqs. (\ref{eq_mom_spin}) and (\ref{eq_mom_chan}).
The full moment is normalized to unity.
}
\label{fig_phase}
\end{center}
\end{figure}

The phase diagram of the 2ch KL near $n_{\rm c}=2$ (half filling) is shown in 
FIG. \ref{fig_phase} where $n_{\rm c}$ is a number of conduction electrons per site.
Figure \ref{fig_phase}(a) shows the transition temperature of antiferro(AF)-spin order that has been discussed in Ref. \cite{jarrell97}.
With lower transition temperature at half-filling,
a ferro(F)-channel order also emerges as shown in FIG.\ref{fig_phase}(b),  
as has been pointed out in Ref. \cite{nourafkan08}.
In the present case, the F-channel phase is masked by the AF-spin phase around half-filling because of the lower transition temperature. 
In real systems, however, one may encounter possible
suppression of the AF-spin order caused, e.g., by 
geometrical frustration, or substantial next-nearest neighbor hopping.
In such a case, the F-channel order may be stabilized.
Because of its unique and interesting property to be described below, 
this letter concentrates mainly on the F-channel phase from now on.

Let us define the spin and channel moments:
\begin{align}
m_{\rm spin} = \sum_{\alpha} 
\langle n_{\alpha\uparrow}  -  n_{\alpha\downarrow} \rangle,
\label{eq_mom_spin}\\
m_{\rm chan} = \sum_{\sigma} 
\langle n_{1\sigma} 
-  n_{2\sigma} \rangle
, \label{eq_mom_chan}
\end{align}
where $n_{\alpha\sigma}$ is the local number operator
of conduction electrons with channel $\alpha$ and spin $\sigma$.
The lower panels of FIG. \ref{fig_phase} show calculated results of the moments.
The spin moment at $n_{\rm c} = 2$ in AF-spin phase becomes maximum with highest transition temperature $T_{\rm spin}$, and gradually decreases away from half filling.
In the case of F-channel phase, on the contrary, the channel moment 
$m_{\rm chan}$ becomes finite only away from half filling, and remains tiny. 
Even though the transition temperature $T_{\rm chan}$ takes the maximum at $n_{\rm c} = 2$, we observe $m_{\rm chan}=0$.
Hence, the channel moment is not a proper order parameter. 

\begin{figure}
\begin{center}
\includegraphics[width=85mm]{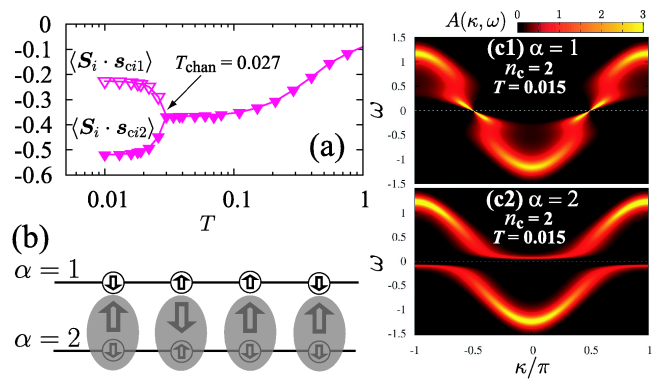}
\caption{
(color online)
(a) Temperature dependence of local correlation functions, and (b) schematic picture of the F-channel phase.
The arrows on the thin lines show conduction electrons, and the shaded ovals show the Kondo singlets centered on each lattice site.
(c) Single-particle spectra of conduction electrons with $\alpha = 1$ and $2$ in this phase are shown in (c1) and (c2), respectively.
}
\label{fig_picture}
\end{center}
\end{figure}

Let us identify the proper order parameter in the F-channel phase.
It has been found that the double occupancy 
$\langle n_{\alpha\uparrow}n_{\alpha\downarrow}\rangle$
in the ordered phase becomes different between $\alpha=1$ and $\alpha=2$ 
\cite{nourafkan08}. 
We propose, however, that the fundamental order parameter involves the localized spin, and hence the Kondo effect.  
The small difference in double occupation comes out as a consequence of the fundamental order parameter.
We shall demonstrate that the proper order parameter leads to identifying an odd-frequency order.  As shown in FIG. \ref{fig_picture}(a), 
the local spin correlations 
$\langle \bm{S}_i\cdot \bm{s}_{{\rm c}i \alpha} \rangle$ becomes different 
below $T_{\rm chan}$.
Namely each localized spin tends to form the Kondo singlet selectively with one of the two conduction bands.
The order parameter is hence given by 
\begin{align}
\Psi \equiv
\langle \bm{S}_i \cdot (\bm{s}_{{\rm c}i1} - \bm{s}_{{\rm c}i2}) \rangle,
\label{Psi}
\end{align}
which is independent of site index $i$.
Since $\Psi$ grows continuously below the transition temperature, 
the phase transition is of second order. 
Note that the order parameter $\Psi$ is a composite quantity, and cannot be described by a one-body mean field such as 
$\langle \bm S_i\rangle$ or 
$\langle 
\bm{s}_{{\rm c}i1} - \bm{s}_{{\rm c}i2} 
\rangle$.

Real-space image of the electronic state is illustrated in FIG. \ref{fig_picture}(b).  For channel $\alpha=1$, the effective Kondo coupling tends to zero, while for $\alpha=2$ the coupling tends to infinity.  Thus the F-channel phase is the mixture of
weak- and strong-coupling limits depending on channels.  This state therefore cannot be accessible by perturbation theory from either limit.

The peculiar character of the F-channel phase appears also in the single-particle spectrum. 
We have derived the single-particle spectrum explicitly
from the imaginary part of the Green function.
Since the self-energy is local in the DMFT, the wave-vector enters only through $\varepsilon_{\bm{k}}$.
We introduce the parameter $\kappa$ defined by $\varepsilon_{\bm{k}} = -D \cos \kappa$, and visualize the spectrum as if the system were in one dimension.
Accordingly the single-particle spectrum is written as $A(\kappa, \omega)$.
Figures \ref{fig_picture}(c1) and (c2) show the spectra of conduction electrons with $\alpha = 1$ and $\alpha =2$, respectively.
The spectrum of the channel $\alpha=1$ 
displays the Fermi-liquid behavior.
Here the mass enhancement factor is estimated as $m^*/m \sim 1.95$ from analysis of the self-energy.
As shown in FIG. \ref{fig_picture}(c2), on the contrary, another channel $\alpha=2$ acquires the insulating character.
The spectrum is almost the same as that of the ordinary Kondo insulator. 
Thus, the F-channel phase consists of a Fermi liquid with $\alpha = 1$ plus Kondo insulator with $\alpha = 2$.
Hence the phase transition at $T=T_{\rm chan}$ is regarded as an orbital-selective metal-insulator transition.
The reason for metal and insulating characters of each channel is understood easily from 
FIG. \ref{fig_picture}(b).

We proceed to show that the F-channel ordered state can be interpreted as an odd-frequency order, which is a consequence of the composite nature of $\Psi$ given by (\ref{Psi}).
Let us 
take the case $n_{\rm c} = 2$ with particle-hole symmetry, and
compare the F-channel order to the AF-spin order.
We take the difference of local Green functions 
and introduce the quantities:
\begin{align}
\Delta G_{\rm spin} (\tau) = \sum_{\alpha} [G^{\rm A}_{\alpha\uparrow}(\tau) - G^{\rm A}_{\alpha\downarrow}(\tau)]
, \label{eq_g_spin_tau} \\
\Delta G_{\rm chan} (\tau) = \sum_{\sigma} [G_{1\sigma}(\tau) - G_{2\sigma}(\tau)] 
, \label{eq_g_chan_tau} 
\end{align}
where 
$G^i_{\alpha\sigma} (\tau) = - \langle T_\tau c_{i\alpha\sigma} (\tau) c_{i\alpha\sigma}^\dagger \rangle$.
In the AF-spin phase, 
$m_{\rm spin} = \Delta G_{\rm spin} (0)$
becomes nonzero, and serves as the order parameter. 
In the F-channel phase, on the other hand, 
we have seen in FIG.\ref{fig_phase}(b) that 
the apparent order parameter 
$m_{\rm chan} = \Delta G_{\rm chan} (0)$ is tiny and vanishes at $n_{\rm c}=2$.

\begin{figure}
\begin{center}
\includegraphics[width=85mm]{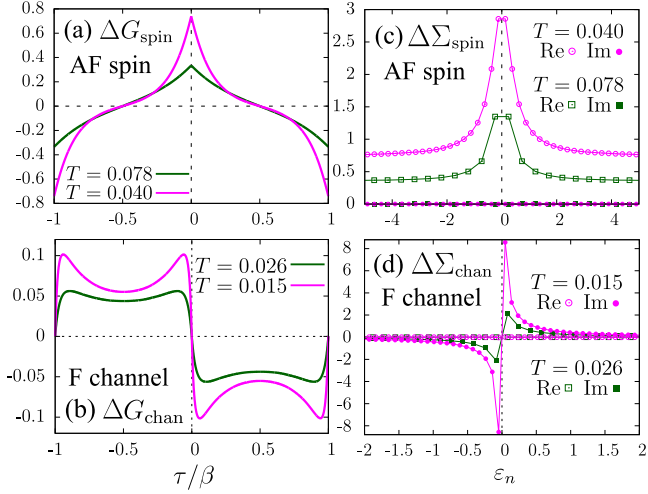}
\caption{
(color online)
(a) Spin- and (b) channel-dependent differences of local Green functions defined in eqs. (\ref{eq_g_spin_tau}) and (\ref{eq_g_chan_tau}).
The differences of complex self energies defined in eqs. (\ref{eq_self_spin}) and (\ref{eq_self_chan}) are shown in (c) and (d).
}
\label{fig_green}
\end{center}
\end{figure}

Figure \ref{fig_green} (a) and (b) show the $\tau$-dependence of these quantities for the case of $n_{\rm c}=2$.
It can be seen that $\Delta G_{\rm spin} (\tau)$ in the AF-spin phase
is an even function of $\tau$, while
$\Delta G_{\rm chan} (\tau)$ in the F-channel phase
is an odd function.
Hence the derivative $\partial\Delta G_{\rm chan}/\partial \tau$ at $\tau = 0$, rather than $m_{\rm chan} 
$, is the proper quantity to describe
the F-channel order.
The explicit form is given by
\begin{eqnarray}
\left. \frac{\partial \Delta G_{\rm chan}}{\partial \tau} \right|_{\tau=0} &=&
\frac{1}{N} \sum _{\bm{k} \sigma} \varepsilon _{\bm{k}} 
\langle c_{\bm{k}1 \sigma}^\dagger c_{\bm{k}1 \sigma} -  c_{\bm{k}2 \sigma}^\dagger c_{\bm{k}2 \sigma} \rangle  \nonumber \\
&&+ \frac{J}{N} \sum_{i} \langle \bm{S}_i \cdot (\bm{s}_{{\rm c} i 1} - \bm{s}_{{\rm c} i 2}) \rangle
, \label{eq_deriv}
\end{eqnarray}
where $N$ is the number of sites.
The second term in the right-hand side (RHS) corresponds precisely to the order parameter $\Psi$ introduced by (\ref{Psi}).
On the other hand, the kinetic energy given by the first term also becomes dependent on channel.
However, this quantity is not suitable as a proper order parameter, since it does not involve localized spins.

The low-energy limit of the self-energy corresponds to a one-body mean field at the Fermi level. Then we introduce
the difference of self-energies as 
\begin{align}
\Delta\Sigma_{\rm spin} (\imu \varepsilon_n)
 = \sum_{\alpha} [\Sigma^{\rm A}_{\alpha\uparrow}(\imu \varepsilon_n) - \Sigma^{\rm A}_{\alpha\downarrow}(\imu \varepsilon_n)]
, \label{eq_self_spin} \\ 
\Delta\Sigma_{\rm chan} (\imu \varepsilon_n)
 = \sum_{\sigma} [\Sigma_{1\sigma} (\imu \varepsilon_n) - \Sigma_{2\sigma} (\imu \varepsilon_n)]
, \label{eq_self_chan}
\end{align}
where $\varepsilon_n = (2n+1)\pi T$ is the fermionic Matsubara frequency.
Figure \ref{fig_green} (c) and (d) show the computed results for 
$\Delta\Sigma (\imu \varepsilon_n)$.
For AF-spin phase,
$\Delta\Sigma_{\rm spin}$ is real and even function of Matsubara frequency $\varepsilon_n$.
For F-channel phase, on the other hand, $\Delta\Sigma_{\rm chan}$ 
is pure imaginary and odd function of $\varepsilon_n$.
Hence, the F-channel order is identified as an odd-frequency order also from the self-energy.
It should be noted that the present clear-cut classification applies only to systems with the particle-hole symmetry.
Otherwise, both orders involve finite components of the opposite parity, and are neither completely even nor odd.
We emphasize that the concept of the composite order is valid even without the particle-hole symmetry.

Let us now discuss the AF-spin and F-channel orders 
in terms of susceptibilities.
We first introduce the two-particle Green function
$\chi_{\bm{q}\sigma \sigma'}^{\alpha\alpha'} (\imu \varepsilon_n , \imu \varepsilon_{n'} ; \imu \nu_m)$
where 
$\nu_m = 2m\pi T$ is an bosonic Matsubara frequency.
The AF-spin and F-channel orders correspond to 
wave vectors $\bm{q} = \bm{Q}$ and $\bm{q} = \bm{0}$, respectively.
The relevant two-particle Green functions for our purpose
are obtained by appropriate linear combinations with respect to channel and spin indices in
$\chi_{\bm{q}\sigma \sigma'}^{\alpha\alpha'}$, and are written as 
$\chi_{\rm spin}^{\rm AF} (\imu \varepsilon_n , \imu \varepsilon_{n'} ; \imu \nu_m)$ and 
$\chi_{\rm chan}^{\rm F} (\imu \varepsilon_n , \imu \varepsilon_{n'} ; \imu \nu_m)$.
Then the following even- and odd-frequency susceptibilities are obtained:
\begin{align}
\chi_{\rm even} &= \frac{1}{\beta} \sum_{nn'} \chi (\imu \varepsilon_n, \imu \varepsilon_{n'}; 0)
, \label{eq_even} \\
\chi_{\rm odd} &= \frac{1}{\beta} \sum_{nn'} g(\varepsilon_n) g(\varepsilon_{n'}) \chi (\imu \varepsilon_n, \imu \varepsilon_{n'}; 0), \label{eq_odd}
\end{align}
where we have suppressed the indices such as AF and spin.
The even part $\chi_{\rm even}$
coincides with the conventional susceptibility, and describes fluctuations of spin and channel moments.
The odd function $g(\varepsilon_n)$ in eq. (\ref{eq_odd}) abstracts the odd-frequency part of the two-particle Green function \cite{jarrell97,sakai04}.

In order to determine the appropriate $g(\varepsilon_n)$,
we introduce an operator ${\cal O}$ by \cite{anders02}
\begin{align}
 {\cal O} = \frac{1}{N} \sum_{i\sigma} \left[
 c_{i1\sigma}^\dagger \frac{\partial c_{i1\sigma}(\tau)}{\partial \tau}
-c_{i2\sigma}^\dagger \frac{\partial c_{i2\sigma}(\tau)}{\partial \tau}
\right]_{\tau=0}
.
\end{align}
The expectation value $\langle {\cal O} \rangle$ coincides with the derivative (\ref{eq_deriv}).
With use of ${\cal O}$, one can prove the following relation:
\begin{eqnarray}
&&
\hspace{-10mm}
N 
\int_0^\beta \langle {\cal O}(\tau){\cal O}^\dagger \rangle \diff \tau = 
\nonumber \\
&&
\hspace{-10mm}
- \frac{1}{\beta} \sum_{nn'}  \varepsilon_n  \varepsilon_{n'} \, \chi_{\rm chan}^{\rm F} (\imu \varepsilon_n, \imu \varepsilon_{n'}; 0) e^{\imu \varepsilon_n 0^+}e^{\imu \varepsilon_{n'} 0^+}
\hspace{-1mm} + \hspace{-1mm} \frac 2N \langle {\cal H} \rangle
, \label{eq_sus2}
\end{eqnarray}
where convergence factors enter in frequency summation.
The first term in the RHS must diverge at the phase transition.
It can be shown that replacing $\varepsilon_n$ by $g(\varepsilon_n)$ leads to 
the divergence at the same transition temperature,
provided $g(\varepsilon_n)$ has finite overlap with $\varepsilon_n$ as a vector in frequency space.
For simpler calculation, we want to avoid the convergence factor, and 
choose $g(\varepsilon_n) = \tanh (\varepsilon_n/D)$ in eq. (\ref{eq_odd}).  
Because of the minus sign in the first term of the RHS of (\ref{eq_sus2}), 
$\chi_{\rm odd}$ becomes negative infinity at the transition temperature.

\begin{figure}
\begin{center}
\includegraphics[width=70mm]{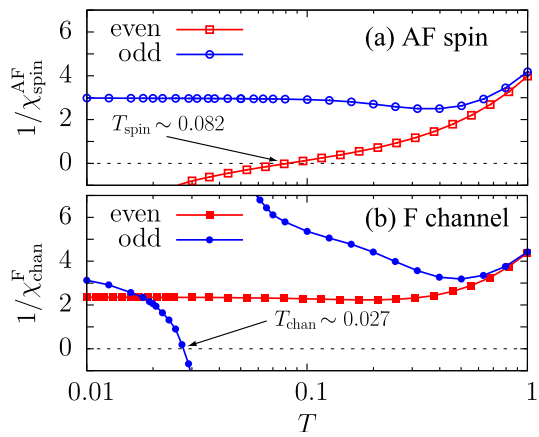}
\caption{
(color online)
Temperature dependence of 
$1/\chi_{\rm even} $ and $1/\chi_{\rm odd}$ for 
(a) AF-spin and (b) F-channel orders.
Phase transitions are signaled by the zero crossing of inverse susceptibilities.
}
\label{fig_suscep}
\end{center}
\end{figure}

Figure \ref{fig_suscep} shows temperature dependence of inverse susceptibilities
in the particle-hole symmetric case $n_{\rm c}=2$.
For the AF-spin order as shown in FIG. \ref{fig_suscep} (a), $\chi_{\rm even}$
diverges at $T_{\rm spin} \simeq 0.082$ 
indicating the spin order below $T_{\rm spin}$, while $\chi_{\rm odd}$
does not show any anomaly.
On the other hand, for the F-channel order as shown in FIG. \ref{fig_suscep} (b), 
$\chi_{\rm even}$ does not signal any instability.
This corresponds to $m_{\rm chan}=0$ in the F-channel phase.
However, $\chi_{\rm odd}$ shows drastic temperature dependence.
With decreasing temperature, $1/\chi_{\rm odd}$ becomes zero
at $T_{\rm chan} \simeq 0.027$ from negative side.
This means that $\chi_{\rm odd}$ 
defined by (\ref{eq_odd}) indeed signals the divergent fluctuation of composite order parameter $\Psi$.

Finally, we discuss possible relevance of our results in understanding real $f$-electron systems.   If we interpret the pseudo-spin as describing a non-Kramers doublet, 
the channel index $\alpha=1$ is translated into the real up spin, and
$\alpha=2$ into the down spin.  Here, the choice of up and down spins can of course be reversed.  
Then the AF spin phase corresponds to AF orbital order, while
the breakdown of channel symmetry is translated into that of the real-spin symmetry.
Namely, the time reversal symmetry is spontaneously broken in F-channel phase of the non-Kramers doublet lattice.  Without the particle-hole symmetry,  the present order gives rise to a tiny magnetic moment corresponding to $m_{\rm chan}$ in FIG.\ref{fig_phase} (b).   However,  the proper order parameter is given by $\Psi$, which actually represents an octupole moment involving conduction electrons.  
Namely, spatial extent of up and down spins become different from each other, 
since conduction electrons with up spin forms a Fermi liquid, while those with down spin tend to form orbital Kondo singlets with localized non-Kramers doublets.   
The resultant spin imbalance sums up to zero, but gives rise to extended magnetic octupoles centered at each site.

Since $\Psi$ is independent of sites, any diffraction measurement cannot detect the present order.
Nevertheless, transport properties signal clear phase transition involving metal-insulator transition of down spins.  
The ordered phase is a kind of half-metal since down spins are insulating.
Moreover, the size of the specific heat anomaly should be substantial since $\Psi$ is of the order of unity as shown in FIG.\ref{fig_picture}(a).
From these characteristics, it is tempting to consider $\Psi$ as a candidate of the hidden order parameter in URu$_2$Si$_2$.  
Of course, much more refinement of the model is necessary before we can push this suggestion further.

In conclusion, we have demonstrated a novel type of symmetry breaking between equivalent channels in the 2ch KL.  The ordered state is characterized by a two-body correlation function, or composite order parameter.   In a general case without the particle-hole symmetry in conduction bands, tiny channel moment also arises as a deceptive order parameter. 
The composite order is regarded as an odd-frequency ordering, which is the first explicit example other than superconductivity.

We are grateful to H. Kusunose for fruitful discussions.
This work was partly supported by a Grant-in-Aid for Scientific Research on Innovative Areas ``Heavy Electrons" (No 20102008) of MEXT.
Numerical calculations were partly performed on supercomputer in ISSP, University of Tokyo.

\end{document}